\documentclass[12pt]{article}

\usepackage{cite,epsfig}

\def\e{\begin{equation}}
\def\f{\end{equation}}
\def\%#1{\mbox{\boldmath $#1$}}
\def\=#1{\overline{\overline #1}}

\def\_#1{{\bf #1}}

\def\.{\cdot}

\def\##1{{\bf#1\mit}}

\def\l#1{\label{eq:#1}}
\def\r#1{(\ref{eq:#1})}
\def\am{\left(\begin{array}{c}}
\def\amm{\left(\begin{array}{cc}}
\def\a{\end{array}\right)}

\begin{document}

\title{Electromagnetic field energy density in artificial
microwave materials with negative parameters}

\author{Sergei
Tretyakov}

\date{Radio Laboratory/SMARAD, Helsinki University of Technology,
P.O. Box 3000, FIN-02015 HUT, Finland\\[3mm]\today}

\maketitle

\begin{abstract}

General relations for the stored reactive field energy density
in passive linear  artificial microwave materials
are
established. These relations
account for dispersion and absorption effects in these materials,
and they are valid also in the regions where the real parts of the
material parameters are negative.
These relations always give physically sound positive values for
the energy density in passive metamaterials. The energy density
and field solutions in active metamaterials with non-dispersive
negative parameters are also considered. Basic physical limitations on
the frequency dispersion of material parameters of artificial
passive materials with negative real parts of the effective
parameters are discussed. It is shown that field solutions in
hypothetical materials with negative and non-dispersive parameters
are unstable.

\end{abstract}


\section{Introduction}

During a few recent years, artificial media with negative real
parts of material parameters (called Veselago media, backward-wave media,
double negative media) have
attracted much attention in view of
the first experimental realizations of such media
and because of potential applications in sub-wavelength imaging.
Some other applications have been proposed, including improvement
of antenna performance. In paper \cite{Z}, radiation from a small electric
dipole inside a spherical shell of such material has been considered,
with some conclusions regarding the antenna bandwidth.
A two-dimensional model of an antenna coated by a dispersive
material shell has been considered in
\cite{our}.
These and many other issues involve considerations of the reactive energy
stored in materials with negative parameters,
and the goal of this paper is
to develop a method for
calculation of the stored energy density in complex materials and to
discuss difficulties in the interpretation of the properties of exotic
materials
as well as the validity of the used models of materials with negative
parameters. A general approach that we introduce in this paper allows to
determine the time-averaged energy density
of time-harmonic electromagnetic fields in dispersive and lossy materials
with various dispersion laws.
Also, the possibility to create
artificial materials with dispersion-free and negative parameters using
{\it active} inclusions \cite{motl} is revisited.

It is well known that the field energy density in materials can be
uniquely defined in terms of the effective material
parameters only in case of small (negligible) losses (e.g. \cite{Landau}).
This is because in the general case when absorption cannot be neglected, the terms
\e \_E\.{\partial \_D\over \partial t} +\_H\.{\partial \_B\over \partial t}\f
describe both the rate of changing the stored energy and the absorption rate.
Only if the absorption is negligible, we can write
\e \_E\.{\partial \_D\over \partial t} +\_H\.{\partial \_B\over \partial t}
={\partial w_e\over \partial t}+{\partial w_m\over \partial t},\f
where $w_e$ and $w_m$ are the energy densities of the electric and magnetic
fields, respectively.

For artificial materials based on metal or dielectric inclusions of various
shapes absorption can be neglected when the operational frequency is
far from the resonant frequencies
of the inclusions and from the lattice resonances, if the material is periodical.
For electromagnetic fields whose spectrum is concentrated near
a certain frequency $\omega_0$, the time-averaged energy density in a material with scalar
frequency-dispersive parameters $\epsilon(\omega)$ and $\mu(\omega)$
reads (e.g. \cite{Landau,Jackson})
\e w={1\over 2}\left.{d(\omega\epsilon(\omega))\over{d\omega}}\right|_{\omega=\omega_0}|E|^2
+{1\over 2}\left.{d(\omega\mu(\omega))\over{d\omega}}\right|_{\omega=\omega_0}|H|^2.
\l{Wd}\f
If in the vicinity of the operating frequency $\omega_0$
the frequency dispersion can be
neglected and $\epsilon$ and $\mu$ can be assumed to be independent
from the frequency,  \r{Wd} simplifies to
\e w={1\over 2}\epsilon |E|^2 +{1\over 2}\mu |H|^2.
\l{Wn}\f
The validity of this formula is
restricted to positive values of $\epsilon$ and $\mu$
because no  passive media in thermodynamic equilibrium can
store negative reactive energy, as this is forbidden by the thermodynamics
(the second principle)\footnote{In thermodynamically non-equilibrium states,
e.g. in non-uniform magnetized plasmas,
the field energy may take negative values \cite{Kadomtsev}
leading to power amplification and instabilities. We do not consider such
situations in this paper.}
\cite{Landau,Vainstein}. Actually, this means that frequency dispersion
cannot be neglected when estimating the stored energy in the frequency regions where the material
parameters are negative.

If the material has considerable losses near the frequency of interest,
it is not possible to define the stored energy density in a general
way (more precisely, it is not possible to express that in terms of
the material permittivity and permeability functions) \cite{Landau}.
Knowledge about the material microstructure is necessary to find the
energy density, and this problem is far from trivial. In the literature,
the energy density expressions have been derived for
an absorbing classical dielectric (Lorentz dispersion) with a single
resonant frequency \cite{Loudon} and for the case where also
the permeability obeys the same dispersion law as the permittivity
\cite{Ruppin}. The known artificial materials with negative real parts of the
material parameters have different and more general dispersion laws, which means
that we need to develop a more general approach suitable for
calculations of the stored energy density in general dispersive and lossy
materials. Such method will be presented in this paper.

In the next section we start from a brief review of the general
properties of the constitutive parameters of passive low-loss materials,
when the energy density can be defined in terms of the
effective material parameters by formula \r{Wd}.

\section{Limitations on material parameters and dispersion in low-loss
passive linear media}

In this section we consider passive linear materials
in thermodynamic equilibrium in the frequency regions where absorption can be
neglected and the field energy density can be found in terms of
the effective permittivity and permeability functions. For simplicity of writing,
we restrict the analysis to isotropic media.

\subsection{Limitations on material parameters}
\label{limitations}

Landau and Lifshitz in \cite[\S 84]{Landau} give a proof that for all
linear {\it passive}
materials in the frequency regions with weak absorption the value of $w$ in
\r{Wd} is not only always
positive\footnote{Positiveness of the derivatives in \r{Wd}
is equivalent to the Foster theorem in the circuit theory.}, but
it is always larger than the energy density of the same fields $\_E$ and
$\_H$ in vacuum. Indeed,  the following inequalities
can be derived from the causality
requirement assuming negligible losses \cite{Landau}:
\e {d\epsilon(\omega)\over{d\omega}}> 0 \f
(Eq.~(84.1) in \cite{Landau})
and
\e {d\epsilon(\omega)\over{d\omega}}> {2(\epsilon_0-\epsilon)\over\omega} \l{842} \f
(Eq.~(84.2) in \cite{Landau}).
Summing these two inequalities one finds that \cite[\S 84]{Landau}
\e {d(\omega\epsilon(\omega))\over{d\omega}}>\epsilon_0.
\l{limit} \f
The same is true for the permeability as well:
\e {d(\omega\mu(\omega))\over{d\omega}}>\mu_0 . \l{broken}\f
This has a clear physical meaning: To create fields in a material,
work must be done to polarize the medium, which means that in the absence of
losses more energy will be stored in material than in vacuum.
This result is very general and applies also to passive low-loss metamaterials with
negative parameters.

Inequality \r{842} can be cast in equivalent form
\e {d(\omega\epsilon(\omega))\over{d\omega}}>2\epsilon_0-\epsilon(\omega) . \l{2e}\f
Depending on the value of $\epsilon$, either \r{limit} or
\r{2e} is stronger. As is seen from \r{2e}, when the permittivity is negative and large
in the absolute value, the permittivity must be very dispersive.

Considering plane electromagnetic waves in transparent isotropic
dispersive materials, Sivukhin \cite{Sivukhin} gave one more
limitation on the relative material parameters:
\e {d(\omega\epsilon_r(\omega))\over{d\omega}}+{\mu_r\over \epsilon_r}
{d(\omega\mu_r(\omega))\over{d\omega}}>0. \f
This relation holds if {\it both} $\epsilon_r$ and $\mu_r$ are either
positive or negative.

If in a certain model the material parameters are assumed to be
completely lossless, the above inequalities can become
equalities. For example, the lossless plasma permittivity function
\e \epsilon(\omega)=\epsilon_0\left(1-{\omega_p^2\over\omega^2}\right)\f
is just on the allowed limit, because in this case
\e {d(\omega\epsilon(\omega))\over{d\omega}}=\epsilon_0\left(1
+{\omega_p^2\over\omega^2}\right)=2\epsilon_0-\epsilon(\omega)\f
It is easy to check that the lossless Lorentz permittivity
model
\e \epsilon(\omega)=\epsilon_0\left(1+{\omega_p^2\over{\omega_0^2-\omega^2}}\right)\f
satisfies all the above inequalities at all frequencies.

Modeling of artificial magnetic materials requires more
care because the very notion of permeability
loses its meaning at high frequencies. Thus, the model permeability
expressions obtained from quasi-static considerations do not
necessarily satisfy the basic physical requirements at high frequencies.
An important example is the effective permeability of
a mixture of chiral or omega particles \cite{chiral}
or split-ring resonators \cite{Kostin,Pendry}, or of  arrays of ``swiss rolls"
\cite{Pendry}:
\e \mu=\mu_0\left(1+{A\omega^2\over{\omega_0^2-\omega^2}}\right).
\l{mu_srr} \f
This function has a physically sound behavior at low
frequencies [$\mu(\omega)=O(\omega^2)$] and near the resonance,
but in the limit $\omega\rightarrow \infty$ it does not tend to $\mu_0$
However, in the limit of extremely high frequencies
materials cannot be polarized at all because of inertia of
electrons, so the parameters must tend to
$\epsilon_0$ and $\mu_0$.\footnote{For this reason, some authors use the
Lorentz dispersion law \r{Lor} to model the effective permeability of dense
arrays of split-ring
resonators, e.g. \cite{ss,ss1}. That model is physically sound at high
frequencies, but fails in the low-frequency limit.}
As a result, this expression becomes non-physical
[due to instantaneous response of the material, condition \r{broken} is not satisfied]
at frequencies larger than $\sqrt{3}\omega_0$.
As explained in \cite[\S 82]{Landau}, the integrals in the Kramers-Kr\"onig
relations should be truncated at a high enough frequency where
the permeability becomes nearly real and constant (formula (82.17) in
\cite{Landau}). Note, however, that inequality
\e {d\mu(\omega)\over{d\omega}}> 0 \f
for the permeability function \r{mu_srr}
is still satisfied at all frequencies.


\subsection{Instability of field solutions in media
with non-dispersive negative parameters}
\label{unstable}

In this section we will introduce a circuit model for
complex passive metamaterials needed to study the energy
density in complex media and use it first to show that although
field solutions of the Maxwell equations for materials with
non-dispersive negative material parameters exist, they are unstable.
This fact appears to be quite obvious, but for some reason it appears to be neglected in
the literature. Indeed, let us consider a material model
with constant and negative $\epsilon$ and $\mu$, as e.g. in \cite{Z}.
Within the frame of this model, the energy density is given
by \r{Wn} and it is negative \cite{Z}. If due to a fluctuation the field
at a certain moment of time increases, the stored energy will
{\it decrease}. In this situation the field will increase even more, so that
the total energy of the system be minimized.

As a first simple example,
let us consider a parallel-plate capacitor filled by a
{\it hypothetical} material with
a negative permittivity $\epsilon<0$ which does not depend on the
frequency. Let us assume that the capacitor
is charged and the voltage between its terminals is
$u(0)$. At $t=0$ it is connected to a resistor
(resistance $R$). The voltage satisfies the differential equation
\e u(t)+CR {du(t)\over {dt}}=0,\f
whose solution is
\e u(t)=u(0)e^{-{t\over{RC}}}.\f
Under our assumption, capacitance $C$ is negative,
which means that the capacitor does not discharge but, on the contrary,
charges itself, collecting free charges from the conductor and
resistor.
This is an expected result, because in this process the
total field energy in the system {\it decreases}.

\begin{figure}
\centering
\epsfig{file=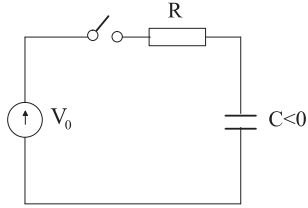, width=0.25\textwidth}
\caption{Connection of a negative-capacitance capacitor to a time-harmonic
voltage source.}
\label{cap}
\end{figure}

The same conclusion follows if the transient regimes in
these systems are analyzed. Figure~\ref{cap} shows a capacitor that is
connected to a time harmonic voltage source $V=V_0\sin(\omega t)$ at time
$t=0$ (there is no charge in the capacitor at $t<0$).
The current in the circuit reads
\e i(t)=V_0\omega C{\cos(\omega t)+\omega RC\sin(\omega t)-e^{-{t\over{RC}}}
\over{1+\omega^2R^2C^2}}.\f
If the capacitance is negative, the current exponentially grows.
This result shows that the time-harmonic solutions
in media with negative {\it non-dispersive} parameters are {\it unstable},
because a small increase in the source amplitude due to noise
will exponentially grow. This is the reason for the know fact \cite{mkk} that
finite-difference time-domain schemes for media with negative and non-dispersive material
parameters are unstable.

In reality, these instabilities do not exist because all passive
materials with negative parameters are frequency dispersive. To illustrate this,
let us consider a more realistic case where the negative permittivity
corresponds to a lossless plasma:
\e \epsilon(\omega)=\epsilon_0\epsilon_r=\epsilon_0
\left(1-{\omega_p^2\over{\omega^2}}\right).\f
At the frequencies $\omega<\omega_p$ the permittivity is real and negative.
The real parameter $\omega_p$ is called {\it plasma frequency.} At microwaves,
dense arrays of ideally conducting wires can be described (although only for specific
excitations \cite{space}) by this
model (e.g. \cite{Brown,Belov,Stas}).

\begin{figure}
\centering
\epsfig{file=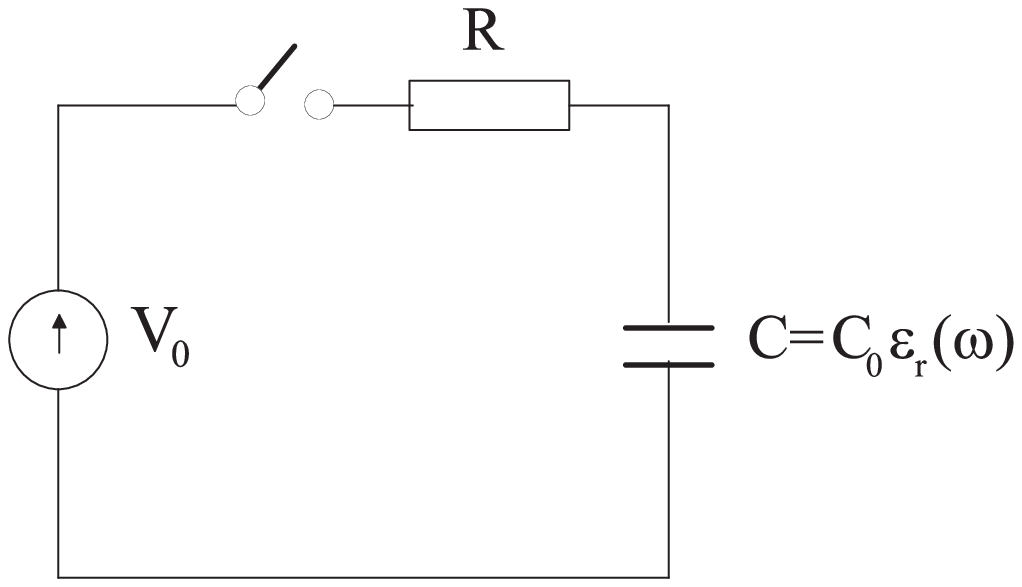, width=0.18\textwidth}\hspace{0.05\textwidth}
\epsfig{file=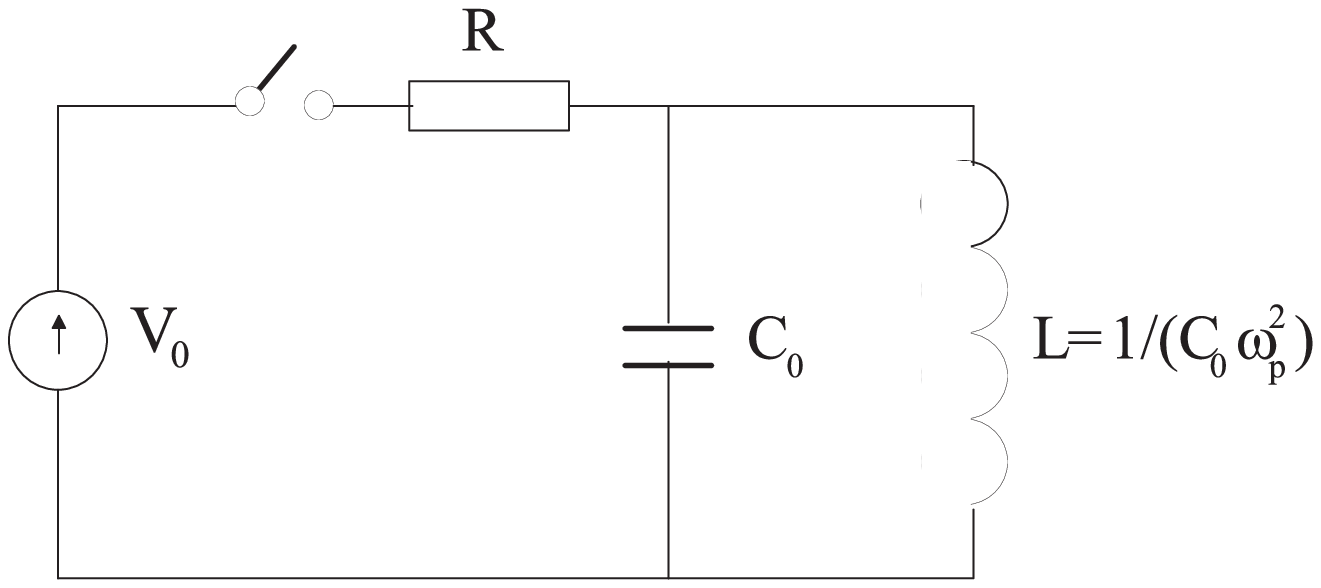, width=0.25\textwidth}
\caption{Connection of a capacitor filled by a lossless plasma to a time-harmonic
voltage source (left) and the equivalent circuit (right).}
\label{cap_dis}
\end{figure}

If we fill a capacitor with such material, its impedance becomes
\e Z={1\over{j\omega C}}={1\over{j\omega C_0 \left(1-{\omega_p^2\over{\omega^2}}
\right) }},  \l{Z_wires}\f
where $C_0$ is the capacitance of the same capacitor filled by vacuum.
Obviously, this corresponds to a parallel connection of a
capacitor with capacitance $C_0$ and an inductor with $L=1/(C_0\omega_p^2)$,
as illustrated in Figure~\ref{cap_dis}.
Connection of a capacitor filled by a lossless plasma to a source is equivalent to
connection of a usual parallel resonant circuit to the same source.
Although the filling material has a negative permittivity at
$\omega<\omega_p$, both the capacitance and inductance in the
equivalent circuit are positive. Naturally, the solution for the
current in this circuit (that can be readily found using the Laplace
transform, for example) does not contain any growing exponents.

Similarly to the case of capacitors filled with {\it hypothetical}
non-dispersive materials with negative permittivities,
current through an inductor filled with a
{\it hypothetical} non-dispersive material with
a negative permeability exponentially grows in time,
if such inductor is connected to a voltage source via a
resistor. For a more realistic consideration,
let us assume that the magnetic material that fills the inductor's coil
can be modeled by the permeability  function \r{mu_srr},
where the magnitude factor $A$ does
not depend on the frequency. In a certain frequency region
(higher than the resonant frequency $\omega_0$) the permeability is
negative.

\begin{figure}
\centering
\epsfig{file=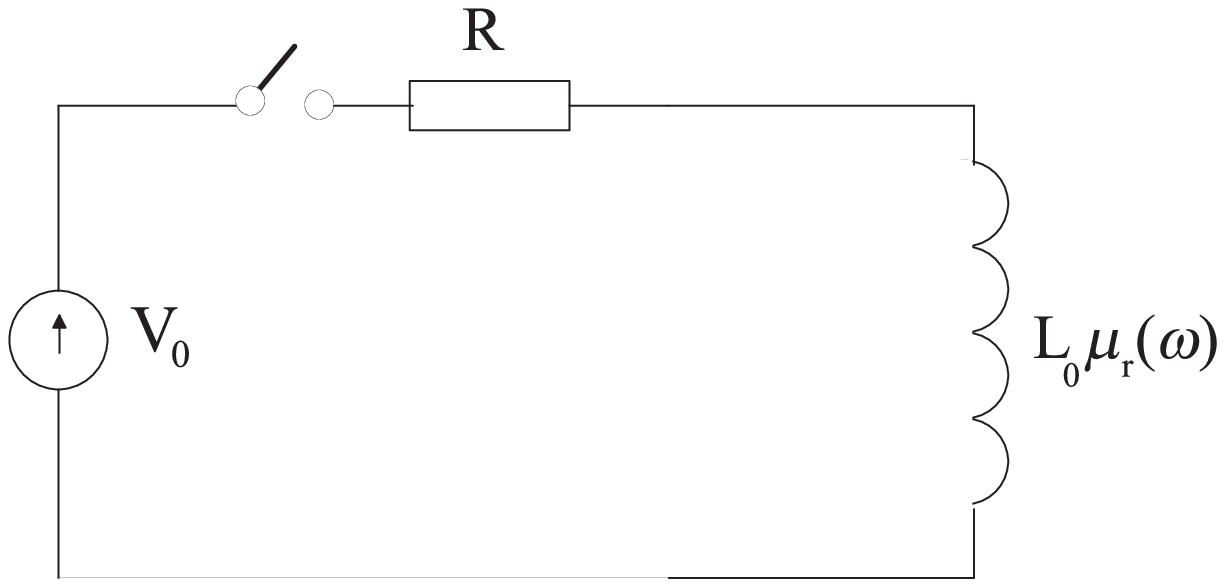, width=0.2\textwidth}\hspace{0.03\textwidth}
\epsfig{file=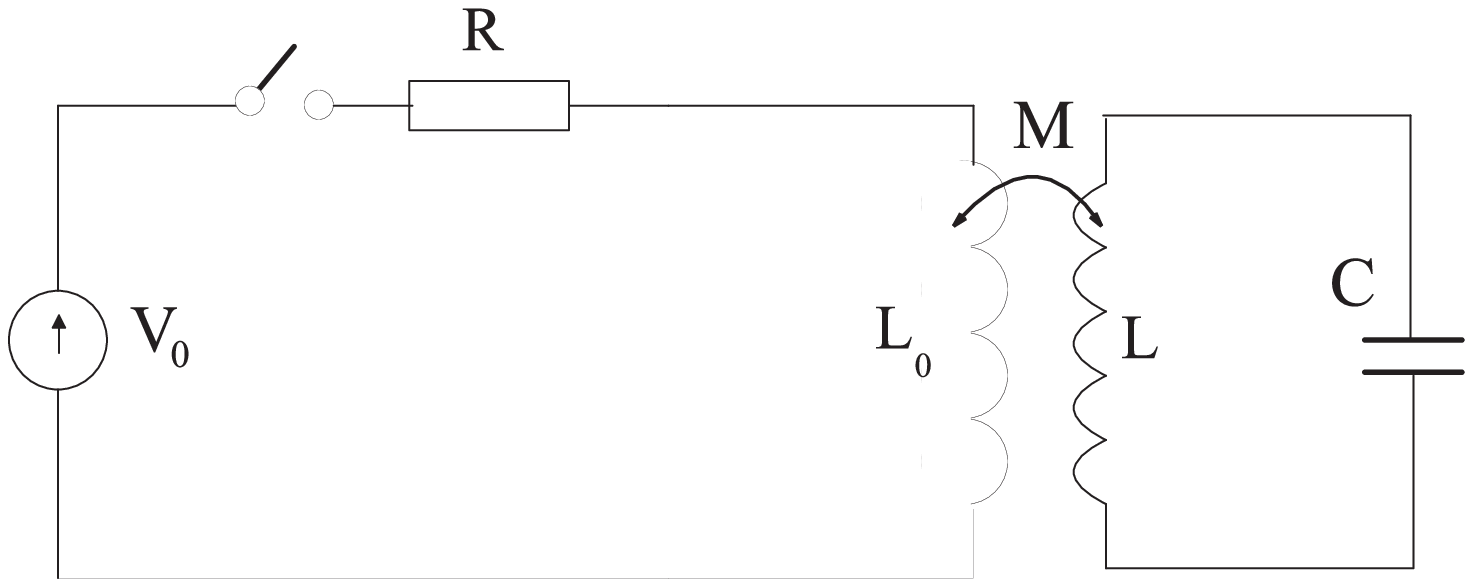, width=0.25\textwidth}
\caption{Connection of an inductor filled by a lossless Lorentzian
magnetic material to a time-harmonic
voltage source (left) and the equivalent circuit (right).}
\label{L}
\end{figure}

Let us consider an inductor filled by this material
and connected via a resistor to a time-harmonic voltage source
(Figure~\ref{L}, left).
The impedance of this inductor
is
\e Z(\omega)=j\omega L_0\mu_r(\omega) = j\omega L_0
+{j\omega^3 L_0 A\over{\omega_0^2-\omega^2}},\f
where $L_0$ is the inductance of the same coil without any magnetic material inside.
This is the same as the impedance of an inductor magnetically coupled to
a resonant circuit (Figure~\ref{L}, right) \cite{our}:
\e Z(\omega)=j\omega L_0
+{j\omega^3 M^2/L\over{\omega_0^2-\omega^2}}.\f
Thus, the circuit shown on the right of Figure~\ref{L}
which contains only usual constant and positive inductances and
a capacitance is equivalent to the circuit
containing a magnetic core with permittivity \r{mu_srr}
(Figure~\ref{L}, left)
if we choose $M^2/L=L_0 A$. As in the case of materials with
negative permittivity, it is obvious that
no instabilities occur in this simple passive circuit.

Next we will determine the
stored energy density in particular realizations of
Veselago media at microwaves, taking into account dispersion and
dissipation.

\section{Passive dispersive and lossy materials with negative parameters}

For media without
magnetoelectric interactions that can be adequately characterized by
two materials parameters: the permittivity and permeability, it is possible
to consider energies stored in the electric and magnetic fields
separately. Indeed, the properties of linear media do not
depend on what particular external field we apply.
Having the full freedom to choose
the external sources, we can always realize a situation where
in a certain (small) volume only electric or magnetic field is non-zero.
Because we deal with {\it effective materials,} the period of the
microstructure or the average distance between inclusions is
considerably smaller than the wavelength, otherwise one cannot introduce
effective permittivity and permeability. Thus, we
can take a representative sample of the material that contains many
inclusions but whose size is still much smaller  than the wavelength, and
probe its properties in (nearly) uniform electric and magnetic fields.

\subsection{Field energy density in wire media}

Negative effective permittivity is most often realized by dense arrays of
parallel thin metal wires. For plane electromagnetic waves whose wave vector is
orthogonal to the wires,  the effective permittivity for electric fields directed
along the wires can be modeled by the plasma permittivity function
\e \epsilon=\epsilon_0\left[1-{\omega_p^2\over{\omega(\omega-j\Gamma)}}\right].
\l{eps_pla}\f
There exist several models for the equivalent plasma frequency $\omega_p$, and here
we will use the quasi-static model \cite{Stas} that is not limited to
the case of small wire radius and allows to estimate the loss factor.
For example, if the skin effect in the wires can be neglected
(uniform current distribution over the wire cross section),
the effective parameters
read \cite{Stas}
\e \omega_p^2={2\pi\over{a^2\epsilon_0\mu_0\log{a^2\over{4r_0(a-r_0)}}}} ,\l{omega_p}\f
\e \Gamma= {2\over{\sigma\mu_0 r_0^2 \log{a^2\over{4r_0(a-r_0)}} }}. \l{Gamma} \f
Here $a$ is the array period, $r_0$ is the wire radius, $\sigma$ is the conductivity of the
wire material, and $\epsilon_0$ and $\mu_0$ are the parameters of the
matrix. The matrix is assumed to be a lossless magnetodielectric, so
$\epsilon_0$ and $\mu_0$ are real numbers.

\begin{figure}
\centering
\epsfig{file=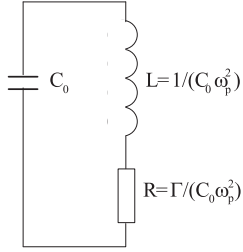, width=0.2\textwidth}
\caption{Equivalent circuit of a capacitor filled
by a wire medium sample with lossy wires.}
\label{para}
\end{figure}

To determine the stored field energy density in this material,
we position a small (in terms of the wavelength or the decay factor
in the effective medium) piece of
this material in a parallel-plate capacitor.
Generalizing formula \r{Z_wires} using \r{eps_pla}, we have for
the admittance
\e Y=j\omega C_0+{1\over{j\omega L +R}}, \f
where $C_0$ is now the capacitance of the capacitor filled with the
matrix material (permittivity $\epsilon_0$), and
\e L=1/(\omega_p^2C_0), \qquad R=\Gamma /(\omega_p^2C_0). \l{param}\f
Obviously, the equivalent circuit is a parallel connection of a
capacitor and an inductor with a loss resistor, Figure~\ref{para}.
This circuit has the same input impedance as the actual
capacitor filled by a material sample.

However, before using this circuit in order to calculate
the stored reactive energy in the medium, we must ensure that
the circuit structure indeed corresponds to the microstructure of the
material under study. It is well known that from the input impedance of a
circuit it is impossible to uniquely determine the circuit structure. In other
words, different circuits can have the same input impedance at all frequencies
(e.g., \cite{Vainstein}).
In the context of this study this means that our circuit model correctly describes the input
impedance of the material-loaded capacitor, but it may fail to properly model
the material microstructure.

In our particular case the material is realized as an array of wires
along the electric field direction (that is, running from one plate
of the capacitor to the other). Apparently, this array of wires possesses
some inductance and resistance connected in series, so we see that
our model indeed corresponds to the microstructure of the medium, and
we can use it.

In the
time-harmonic regime the time-averaged stored reactive energy is
\e W={1\over 2}\left(C_0|V_C|^2+L|I_L|^2\right) ,\f
where $V_C$ is the voltage amplitude on the capacitor and
$I_L$ is the amplitude of the current through the inductor (see the
equivalent circuit in Figure~\ref{para}).
This can be written as
\e W={1\over 2}C_0|V_C|^2\left[1+{L\over{C_0(\omega^2 L^2+R^2)}}\right] .\l{WW}\f
For a parallel-plate capacitor  (the plate area $S$,
the distance between the plates $d$), we have $C_0=\epsilon_0 S/d$
and $V_C=Ed$. The total energy is the energy density $w_e$ multiplied
by the capacitor volume $Sd$:
\e W=w_eSd={1\over 2}{\epsilon_0 S\over d}|E|^2d^2
\left[1+{L\over{C_0(\omega^2 L^2+R^2)}}\right].\l{ww}\f
Thus, the energy density reads
\e w_e={\epsilon_0 \over 2} \left(1+{\omega_p^2\over{\omega^2+\Gamma^2}}\right)|E|^2,
\l{w_wire}\f
where we have substituted the values of the circuit parameters from
\r{param}. Parameters $\omega_p$ and $\Gamma$ are given by
\r{omega_p} and \r{Gamma}, respectively.

If the losses can be neglected ($\Gamma\ll \omega$),
the same result follows from \r{Wd}, since
\e {d(\omega\epsilon(\omega))\over{d\omega}}=\epsilon_0
\left(1+{\omega_p^2\over{\omega^2}}\right) .\f

Let us next consider a wire medium where the matrix is a lossy
dielectric, with the permittivity $\epsilon=\epsilon_0-j\sigma_d/ \omega$,
where $\sigma_d$ is the conductivity of the matrix material.
Following the approach of \cite{Stas}, we find that the
effective permittivity is
\e \epsilon=\epsilon_0\left[1-{\omega_p^2\over{\omega(\omega-j\Gamma)}}\right]
-j{\sigma_d\over  \omega}, \f
where the plasma parameters $\omega_p$ and $\Gamma$ remain the same as for
wires in a lossless matrix
(the physical reason for this is that the negative
permittivity appears due to the inductance of the wire array, and that
inductance does not depend on the dielectric loss).
In the equivalent circuit shown in Figure~\ref{para},
the matrix loss will be reflected by an additional loss
resistance $R_d=d/(\sigma_d S)$ connected in parallel with the capacitor
$C_0$. This means that the expression for the stored energy
density \r{w_wire} does not change if the matrix has
some non-zero conductivity: dielectric losses in the matrix have no
effect on the stored energy density function, while
losses in the wires have a strong effect.

Let us assume next that the matrix has no conductivity, but
there are some magnetic losses: the matrix parameters are $\epsilon_0$ and
$\mu=\mu_0-j\mu''$, where $\epsilon_0$ and $\mu_0$ are  real.
The effective permittivity of the wire medium in this matrix
can be found using formulas of \cite{Stas} with substitution
$L\rightarrow L(1-j\mu''/\mu_0)$ where $L$ is the inductance per unit length of
the wire array in the matrix with the permeability $\mu_0$.
The result is  the same as \r{eps_pla} where the plasma frequency
does not depend on $\mu''$ and is given by \r{omega_p}, but the
loss factor $\Gamma$ is different:
\e \Gamma= {2\over{\sigma\mu_0 r_0^2 \log{a^2\over{4r_0(a-r_0)}} }} +\omega
{\mu''\over \mu_0}. \f
In addition to the loss factor due to resistive wires,
there is a factor measuring magnetic losses in the background medium.
The structure of the equivalent circuit in Figure~\ref{para} does not
change, but the resistance
\e R={\Gamma \over \omega_p^2C_0} \f
is now {\it frequency-dependent}. However, the stored energy density
can be still calculated using formula \r{w_wire},
because the additional impedance is {\it real}.

\subsection{Field energy density in artificial Lorentzian dielectrics}

Negative permittivity can be alternatively realized using
artificial dielectrics with resonant inclusions. Frequency
dispersion in such materials is described by the Lorentz formula
\e
\epsilon=\epsilon_0\left(1+{\omega_p^2\over{\omega_0^2-\omega^2+j\omega
\Gamma}} \right),\l{Lor}\f
which is widely used as a model of
natural materials in solid state physics. We will consider
microwave materials designed as a collection of short metal
needles\footnote{The inclusion shape is actually not critical for
the validity of the model relation \r{Lor}. They can be metal
ellipsoids or spheres, for example.} oriented along the electric
field direction and distributed  in a lossless matrix. If the
length of the needles is much smaller than the wavelength and the
distance between the needles is much larger than the needle length
but much smaller than the wavelength, formula \r{Lor} is a good
estimate for the effective permittivity. Parameters $\omega_p$,
$\omega_0$ and $\Gamma$ can be estimated in terms of the
particle dimensions and the inclusion concentration using the antenna model
of an individual inclusion
\cite{modeboo} and an appropriate mixing rule (e.g., \cite{modeboo,Sihvola}).

Making use of the same approach as above, we
consider a capacitor filled with a material having this dispersion law.
Its admittance is
\e Y=j\omega C_0\epsilon=j\omega C_0 +{j\omega C_0\omega_p^2
\over{\omega_0^2-\omega^2+j\omega \Gamma}}.\f
Apparently, this is the admittance of a parallel connection of a
capacitor $C_0$ and a series resonant circuit with the elements
\e C=C_0\omega_p^2/\omega_0^2, \quad L=1/(\omega_p^2C_0), \quad
R=\Gamma/(\omega_p^2C_0). \l{pl} \f
It differs from the equivalent circuit for wire media
(Figure~\ref{para}) by the additional capacitance $C$ in series with
$L$ and $R$.
This equivalent circuit is a valid model for the microstructure of this
material because currents along needles are modeled by inductance
$L$ and charges at the ends of the needles by capacitance $C$.
The loss is due to non-ideally conducting material of the
needles (the matrix material is assumed to be
lossless), so it is appropriately modeled by resistor $R$ in series
with the inductance.

The stored reactive energy is the sum of the energies stored in
all reactive elements:
\e W=w_eSd={1\over 2}\left(C_0|V_{C_0}|^2+L|I_L|^2+C|V_C|^2\right), \f
where $V_{C_0}$ and $V_C$ are the voltages at the respective elements.
Solving for $I_L$ and $V_C$ and substituting
the equivalent circuit parameters \r{pl} we find
\e w_e={\epsilon_0\over 2}\left[1+{(\omega^2+\omega_0^2)\omega_p^2
\over{(\omega_0^2-\omega^2)^2+\omega^2\Gamma^2}}\right]|E|^2.\f
This result coincides with that obtained earlier in
\cite{Ruppin}, where the motion equation for the
electric polarization was directly solved. For the
case of negligible losses ($\Gamma\rightarrow 0$),
the same result follows from \r{Wd}.

The present method extends to the case of
many resonant frequencies (multi-phase mixtures of inclusions of
several different sizes) by simply adding more parallel $LCR$ branches
to the equivalent circuit.

\subsection{Field energy density in dense arrays of split rings}

Dense arrays of split rings and other similar
structures can be modeled in the quasi-static regime by
the following effective permeability (e.g. \cite{Kostin,Pendry}):
\e \mu=\mu_0\left(1+{A\omega^2\over{\omega_0^2-\omega^2+j\omega \Gamma}}\right),
\l{Lorentz}\f
where the magnitude factor $A$ and the loss factor $\Gamma$ do
not depend on the frequency.
Similarly to the approach introduced above for artificial
dielectrics, we position a small (in terms of the wavelength or the
decay length in the effective medium) sample in the magnetic field
of a solenoid with inductance  $L_0$. The inductance becomes
\e Z(\omega)=j\omega L_0\mu_r(\omega) = j\omega L_0
+{j\omega^3 L_0 A\over{\omega_0^2-\omega^2+j\omega \Gamma}}. \l{ZL}\f

\begin{figure}
\centering
\epsfig{file=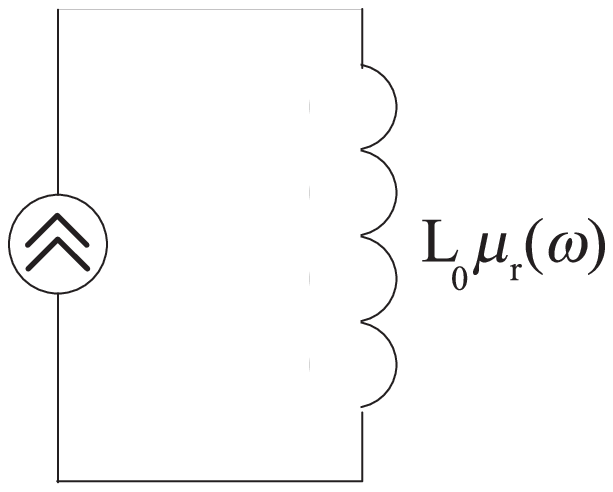, width=0.15\textwidth}\hspace{0.05\textwidth}
\epsfig{file=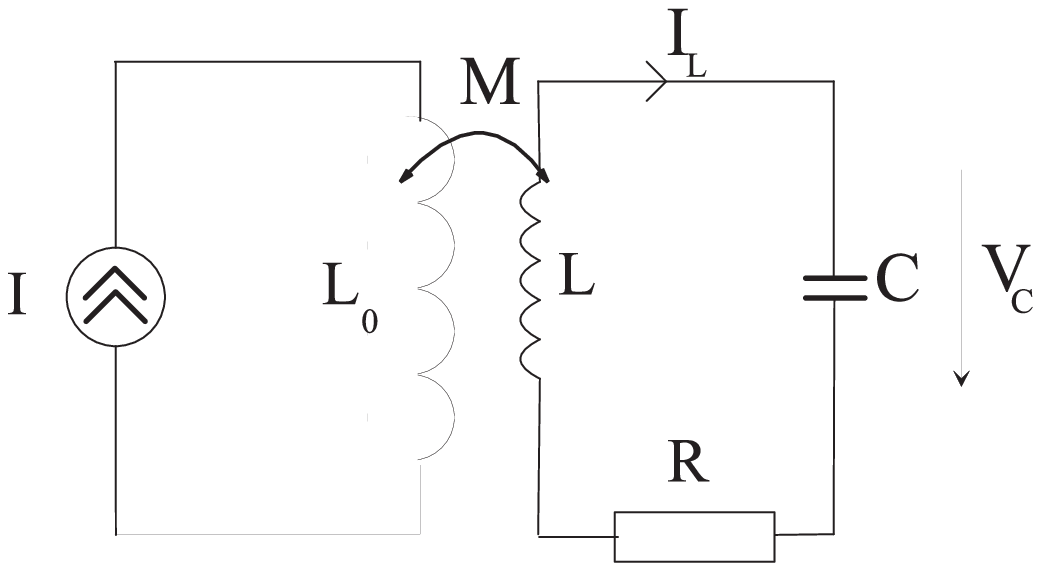, width=0.25\textwidth}
\caption{Magnetic material sample in the probe magnetic
field of a solenoid (left) and the equivalent circuit (right).}
\label{LL}
\end{figure}

An equivalent circuit with the same impedance\footnote{A similar approach
was used in \cite{our}
to determine the equivalent quality factor of a simple radiator
loaded by an artificial magnetic.} is shown in
Figure~\ref{LL}. Indeed, the input impedance seen by the source is
\e Z=j\omega L_0+{j\omega^3 M^2/L\over{{1\over{LC}}-\omega^2
+j\omega {R\over L} }}.\f
This is the same as \r{ZL} if
\e {M^2\over L}=L_0A,\quad {1\over LC}=\omega_0^2,\quad
{R\over L}=\Gamma . \l{Lpara}\f
This is a correct equivalent representation from the microscopic
point of view,
because the material which we model is a collection of
capacitively loaded loops magnetically coupled with the incident
magnetic field.

The total stored reactive energy is the sum of the
energies stored in all reactive elements:
\e W={1\over 2}(L_0|I|^2+  L|I_L|^2 + C|V_C|^2) .\f
Expressing $I_L$ and $V_C$ in terms of $I$, we get, similarly
to the derivations in \cite{our},
\e W={1\over 2}\left[ L_0 +{\omega^2M^2C(1+\omega^2 LC)
\over(1-\omega^2LC)^2 + \omega^2R^2C^2} \right]|I|^2.
\f
Rearranging terms and substituting the equivalent parameters
\r{Lpara}, this can be written as
\e W={1\over 2}L_0|I|^2\left[1+{A\omega^2(\omega_0^2+\omega^2)
\over{(\omega_0^2-\omega^2)^2+\omega^2\Gamma^2}}\right].\f
Considering the stored energy in one unit-length section of the
solenoid, we have
\e W=w_mS={1\over 2}\mu_0 n^2S {|H|^2\over n^2}
\left[1+{A\omega^2(\omega_0^2+\omega^2)
\over{(\omega_0^2-\omega^2)^2+\omega^2\Gamma^2}}\right],\f
where $S$ is the solenoid cross section area and
$n$ is the number of turns per unit length. We have substituted
the solenoid inductance per unit length (a tightly wound long solenoid)
$L_0=\mu_0 n^2 S$ and used the relation $I=H/n$
between the current $I$ and the magnetic field
inside the solenoid $H$.
Finally, the stored field energy density
is found to be
\e w_m={\mu_0\over 2}
\left[1+{A\omega^2(\omega_0^2 +\omega^2)
\over{(\omega_0^2-\omega^2)^2+\omega^2\Gamma^2}}\right]|H|^2.\l{correct_srr}\f

It is important to note that in this particular case
formula \r{Wd} leads to an incorrect expression even if the
losses are negligible ($\Gamma\rightarrow 0$). For $\omega>\sqrt{3}\omega_0$ the stored
energy density obtained from \r{Wd} is less than the energy
stored in vacuum, and it becomes even negative
at still higher frequencies. This is a manifestation of the failure of the
quasistatic permeability model discussed above in Section~\ref{limitations}.
Formula~\r{correct_srr} should be used even in the case of small losses.

\section{Artificial negative-permittivity materials containing active inclusions}

In paper \cite{motl} it was shown that in principle it is possible to
realize a material with {\it non-dispersive} negative material parameters
using artificial ``molecules" that contain electronic circuits (impedance inverters).
In these active materials, the basic physical limitation
\r{limit} does not apply. In the light of the previous consideration, it
is of interest to study if in this case the stored energy density
can be negative and clarify the physical meaning of this effect.

\begin{figure}
\centering
\epsfig{file=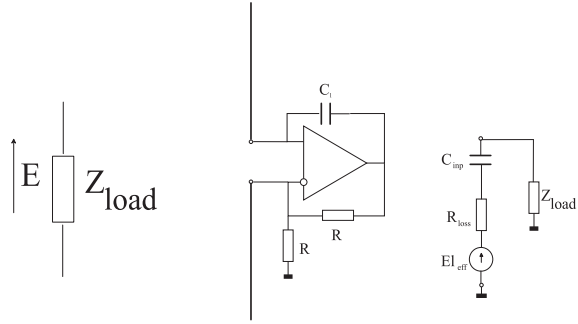, width=0.47\textwidth}
\caption{An artificial molecule in external electric field $E$,
its realization, and its equivalent circuit.}
\label{opam}
\end{figure}

Let us consider an artificial molecule in form of a short dipole antenna
loaded by a bulk impedance (Figure~\ref{opam}, left).
To realize a wide-band non-dispersive negative permittivity, the load
must be a {\it negative} capacitance \cite{motl}:
\e Z_{\rm load}=-{1\over{j\omega C_{\rm inp}}}-
{Nl_{\rm eff}^2\over{j\omega\epsilon_0(1+|\epsilon|)}} ,\l{Zl}\f
where $C_{\rm inp}$ is the input capacitance of the dipole antenna,
$N$ is the number of particles per unit volume, $l_{\rm eff}$ is the
effective antenna length, and $\epsilon$ is the effective negative permittivity that
we want to realize [the low-density limit of the Clausius-Mossotti
formula was used in \cite{motl} to derive \r{Zl}].
On the right of Figure~\ref{opam},
the equivalent circuit of the whole molecule is shown. The negative
capacitance is realized by an impedance inverter, which is connected to
an equivalent circuit of a short dipole antenna ($C_{\rm inp}$ is its input capacitance).
$R_{\rm loss}$ is the loss resistance of the antenna (the radiation
resistance is compensated by the interaction field, as we
suppose that the array of particles is regular or enough dense \cite{modeboo}).

The total impedance connected to the voltage source (external electric field)
reads
\e Z=R_{\rm loss}-{Nl_{\rm eff}^2\over{j\omega\epsilon_0(1+|\epsilon|)}},\f
that is, effectively we have a capacitor filled with a non-dispersive
negative-permittivity material excited by a voltage source. As we know from
section~\ref{unstable}, this system is {\it unstable}, which is
an expected result.
This is in agreement with experiments \cite{private}.

\section{Conclusions}

A general approach that allows to determine the stored energy
density in complex composite microwave materials has been been
presented. The method is based on an equivalent circuit representation of
small material samples excited by electric and magnetic fields.
Introduction of equivalent circuit parameters for specific
microstructures of media is physically equivalent to an appropriate
averaging procedure, needed to determine the properties of the
effective medium.
Particular examples of wire media (negative-epsilon material) and
arrays of split-rings (negative-mu material) have been considered, as well as the usual
Lorentzian dielectrics with losses. The last case has been
considered in the literature using a different approach,
and the present result agrees with the known formula.

The above derivations show how
the energy density can be found for any passive and lossy composite,
if its microstructure is known. The energy density is determined in
terms of the energy stored in the reactive elements of the
equivalent circuits. Naturally, in all cases the stored energy is
positive, as it should be in all passive materials.

This conclusion appears to be very natural if one remembers that
passive metamaterials exhibiting negative material parameters
are anyway made from usual materials like metals or
dielectrics. On the microscopic level, the
stored energy is the electromagnetic field energy in the
matrix material (normally a dielectric) and
in the inclusions (normally a metal of another dielectric). This energy is a
strictly non-negative definite function. The energy stored in a
sample of the effective medium is the average of the corresponding
microscopic quantity, and
there is no reason to expect that for some specific
shapes of metal inclusions the effective material will
store negative energy.
For example, the use of passive metamaterials in the design of antennas
basically means adding some extra metal or dielectric elements
like metal wires or split-ring resonators to a simpler antenna.
On the fundamental level, this means only changing the
antenna shape.

\subsection*{Acknowledgement}

This work has been partially funded by the Academy of Finland and
TEKES through the Center-of-Excellence program.
Helpful discussions with Prof.\ I.S. Nefedov, Dr.\ S.I. Maslovski, and
Prof.\ C.R. Simovski are very much appreciated.

\end{document}